\documentclass[preprint,12pt]{elsarticle}

\usepackage{graphicx}
\usepackage{xcolor}
\usepackage{bm}
\usepackage{amssymb}
\usepackage{soul}

\journal{Physica C: Superconductivity and its Applications}

\begin{document}

\newcommand{\Acknowledgement}[1]{%
    \par\medskip\noindent\textbf{Acknowledgements}\par\medskip
    #1
    \par\medskip
}

\begin{frontmatter}



\title{Describing heat dissipation in the resistive state of three-dimensional superconductors}


\author[inst1]{Leonardo Rodrigues Cadorim}
\ead{leonardo.cadorim@unesp.br}

\author[inst1]{Lucas Veneziani de Toledo}
\ead{lucas.veneziani@unesp.br}

\author[inst1]{Edson Sardella \corref{cor1}}
\ead{edson.sardella@unesp.br}
\cortext[cor1]{Corresponding author}

\affiliation[inst1]{organization={Departamento de Física, Faculdade de Ciências, 
Universidade Estadual Paulista (UNESP)},
            addressline={Av. Eng. Edmundo Carrijo Coube, 14-01}, 
            city={Bauru},
            postcode={17033-360}, 
            state={São Paulo},
            country={Brazil}}



\begin{abstract}
In this work we study the role of the heat diffusion equation 
in simulating the resistive state of superconducting films. By 
analyzing the current-voltage and current-resistance 
characteristic curves for temperatures close to $T_c$ 
and various heat removal scenarios, we demonstrate 
that heat diffusion notably 
influences the behavior of the resistive state, 
specially near the transition to the normal state, 
where heat significantly changes the critical current 
and the calculated resistance. Furthermore, we show how the efficiency of the substrate has important 
effects in the dynamics of the system, 
particularly for lower temperatures. 
Finally, we investigate the hysteresis loops, 
the role of the film thickness and of the 
Ginzburg-Landau parameter, the 
findings reassuring the significance of 
accounting for heat diffusion in accurately 
modeling the resistive state of 
superconducting films and provide valuable 
insights into its complex dynamics. 
To accomplish these findings, we have used 
the $3D$ generalized Ginzburg-Landau equation 
coupled with the heat diffusion equation.
\end{abstract}

\begin{graphicalabstract}
Four sequential frames depicting 
the evolution of a vortex-antivortex pair 
in a superconductor, from nucleation at 
the borders to annihilation at the 
center. The pair traverses the 
$yz$ plane, intersecting both the 
lower and upper $xy$ planes. The left and right columns display the color maps for the order parameter and temperature, respectively. It can be observed that as the vortex-antivortex penetrates deeper into the superconductor, the temperature correspondingly increases. \\[0.5cm] 
\includegraphics[width=\textwidth]{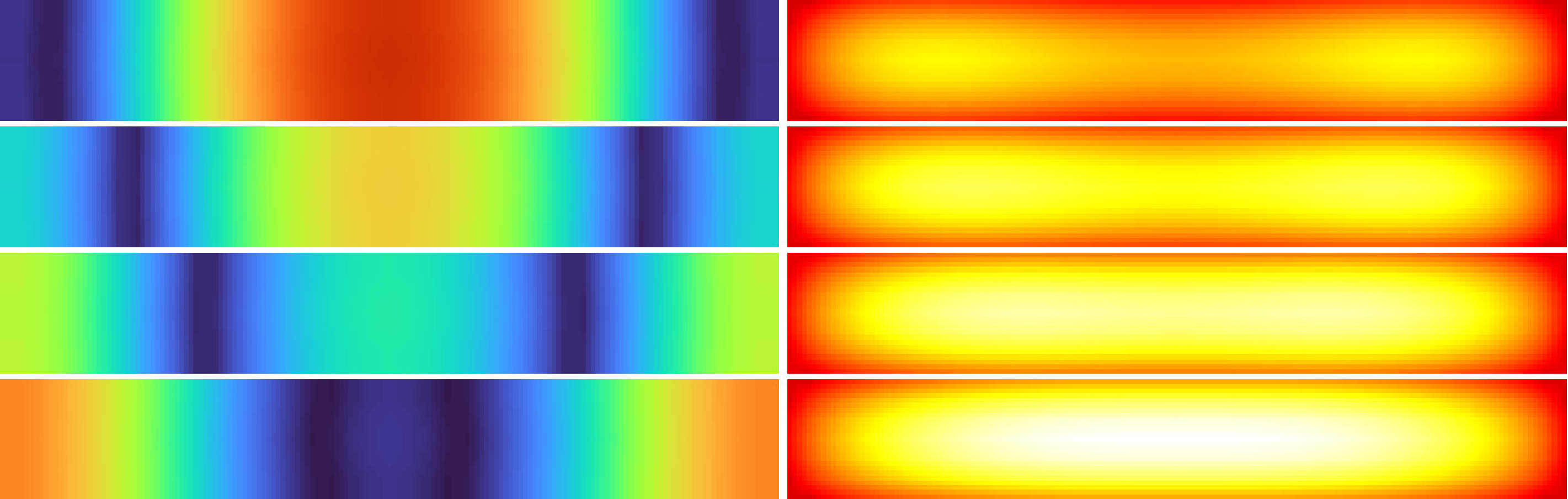}
\end{graphicalabstract}

\begin{highlights}
\item It is introduced a numerical method that enables the study of heat diffusion in three-dimensional superconductors. 
\item It is investigated how heat removal efficiency affects the critical properties of the resistive state of three-dimensional superconductors.
\item It is shown how parameters such as the film thickness and the Ginzburg-Landau parameter affects the heat diffusion process.
\end{highlights}


\end{frontmatter}


\section{\label{sec:sec1}Introduction}


Throughout the years, superconductivity has remained a highly significant and influential subject within the field of condensed matter physics. The study of superconducting phenomena continues to yield insights and practical applications. One intriguing aspect of superconductivity pertains to the resistive state that emerges under certain conditions. Specifically, the self-field of the transport current in superconductors can lead to a resistive state \cite{berdiyorov2009kinematic,berdiyorov2014dynamics,andronov1993,Sivakov2003,cadorim2020ultra}. In type-II superconductors, the circularly shaped magnetic self-field induced by a current gives rise to Abrikosov vortices with ring-like shapes, commonly referred to as closed vortices \cite{cadorim2023}.

In recent decades, the advancement of superconducting devices has been facilitated by the evolution in experimental techniques. These devices are frequently operated by applying currents, a practice that has prompted an intensified analysis of superconducting samples under such conditions \cite{berdiyorov2009kinematic,berdiyorov2014dynamics,Sivakov2003,vodolazov2007rearrangement, presotto2020dynamical}. Among the pivotal considerations in this domain, the interplay of heat diffusion and the superconducting properties of these systems stands out. As the devices progressively shrink in size, their susceptibility to thermal effects also amplifies. The current flow through these superconducting structures has the potential to induce resistive heating due to the intrinsic electrical resistance of materials, even when they are in the superconducting state. This resistive heating can give rise to the generation of heat within the device. Thus, the heat diffusion properties inherent in these nanoscale superconducting systems 
take an undeniable importance to the very functioning of the device itself \cite{lyu2021superconducting, vodolazov2005, aguirre2021vortex}.

In Ref.~\cite{lyu2021superconducting}, for instance, the authors explore the Joule heating produced by the motion of vortices through conformal-mapped nanoholes. The asymmetric distribution of nanoholes causes a vortex motion dependent on the polarity of the current. The disparity in the dissipated heat between the two current directions gives rise to a superconducting diode. It is thus clear that a correct description of heat dispersion becomes indispensable for the modeling and perfection of such devices.

In Ref.~\cite{vodolazov2005} the authors applied a theoretical analysis employing a numerical solution that encompassed the coupled time-dependent Ginzburg-Landau and heat dissipation equations. This  analysis showed a relationship between critical currents and the applied magnetic field within a mesoscopic square structure outfitted with attached contacts. In line with experimental findings, the study elucidated the presence of hysteresis phenomena. These hysteresis effects were ascribed to the substantial heat dissipation occurring within the sample, particularly at current levels that approached the depairing Ginzburg-Landau threshold or were influenced by the dynamic behavior of the superconducting condensate.      

In a previous work of Ref.~\cite{cadorim2023} the authors 
investigated the resitive state of a long superconducting 
film. We describe a crossover phenomenon where straight 
vortex-antivortex (\textit{v-av}) pairs transition to 
curved (closed) \textit{v-av} pairs as the thickness of the 
superconducting film increases. We establish a criteria 
for this transition based on the aspect ratio of the 
vortex and show that it correlates with changes 
in the \textit{IV} curve. An indirect method for detecting closed 
vortices experimentally has been proposed. This method 
involves measuring the time-averaged magnetic flux at 
the lateral sides of the film.
In this paper we will extend this work, now investigating how the three-dimensional dynamics of the closed \textit{v-av} pairs affects the heat diffusion of the superconducting system.

Therefore, the main objective of this study is to investigate the impact of the heat diffusion properties on the resistive state of mesoscopic superconductors with finite thickness, within a fully three-dimensional model. In particular, we aim to explore how the behavior of superconducting samples, with varying thicknesses and applied currents, is influenced by different substrate properties. To achieve this, we consider the effect of temperature fluctuations on the resistive state, thus providing a view of how substrate properties and temperature interplay in the behavior of mesoscopic superconductors.

Moreover, in the above above cited works, the heat diffusion process was studied within a two-dimensional framework, which it is the common practice in the literature. While the $2D$ model is capable of capturing the qualitative features of the vortex dynamics, we show here that critical parameters, such as the critical currents to the onset of the resistive and normal states, obtained by the $2D$ model significantly differs fro the ones obtained within the fully $3D$ model developed in this work. These findings are relevant to future simulations aiming the proposal of novel superconducting devices, where the precise knowledge of the critical parameters of the system are of great importance.

The outline of this paper is as follows. In Section \ref{sec:sec2} 
we present a comprehensive overview of the theoretical 
model, encompassing the generalized 
Ginzburg-Landau equation coupled 
with Ampere's law and heat diffusion equation 
which are essential to describe 
the resistive state of superconductors. 
In Section \ref{sec:sec3}, we detail the  
methodology, covering the employed parameters 
and the numerical techniques utilized to solve these 
equations. The remaining of this Section 
is devoted to presenting both the results 
and discussion. In Section \ref{sec:sec4} we present 
our concluding remarks.

\section{\label{sec:sec2}The Theoretical Model}

To investigate the evolution of superconductivity 
in a dirty superconductor, we use the framework of the generalized 
time-dependent Ginzburg-Landau (GTDGL) equation, 
\cite{kramer1978theory,watts1981nonequilibrium} 
coupled with the Ampère's law. The equations are written as:
\begin{eqnarray}
    \frac{u}{\sqrt{1+\gamma^2|\psi|^2}}\left [ \frac{\partial }{\partial t}+\mbox{i}\varphi
    +\frac{1}{2}\gamma^2\frac{\partial |\psi|^2}{\partial t} \right ] \psi = & & \nonumber \\ 
    = -\left(-\mbox{\mbox{i}\boldmath $\nabla$}-\textbf{A}\right)^{2}\psi+\psi(1-T-|\psi|^2),& & 
    \label{eq:eq1}
\end{eqnarray}
\begin{equation}
    \sigma \left(\frac{\partial \textbf{A}}{\partial t}+\mbox{\boldmath $\nabla$}\varphi\right) 
    = \textbf{J}_s-\kappa^2\mbox{\boldmath $\nabla$}\times\textbf{h},
    \label{eq:eq2}
\end{equation}
where 
\begin{equation}
\textbf{J}_s = {\rm Re}\left [\bar{\psi}{(-\mbox{\mbox{i}\boldmath $\nabla$}-\textbf{A})}\psi\right]
\label{eq:eq3}
\end{equation}
is the superconducting current density.

To obtain the scalar potential in each time 
instant, we use the equation for the continuity of 
electric charge:
\begin{equation}
    \frac{\partial\rho}{\partial t}+\mbox{\boldmath $\nabla$}\cdot\textbf{J} = 0
    \label{eq:eq15}
\end{equation}
where $\rho$ is the electric charge density and the total current $\textbf{J}$ can be decomposed as 
the sum of the superconducting current $\textbf{J}_s$ and 
the normal current $\textbf{J}_n = \sigma\textbf{E}$, with the 
electric field being written as:
\begin{equation}
    \textbf{E} = -\frac{\partial \textbf{A}}{\partial t}-\mbox{\boldmath $\nabla$}\varphi.
    \label{eq:eq16}
\end{equation}

Replacing Eq.~(\ref{eq:eq16}) into Eq.~(\ref{eq:eq15}) and assuming 
there is no charge accumulation, that is $\partial \rho/\partial t = 0$, we obtain:
\begin{equation}
    \nabla^2\varphi = \frac{1}{\sigma}\mbox{\boldmath $\nabla$}\cdot\textbf{J}_s,
    \label{eq:eq20}
\end{equation}
where we have also adopted the Coulomb gauge $\mbox{\boldmath $\nabla$}\cdot\textbf{A} = 0$.

Here, Eqs.~(\ref{eq:eq1}-\ref{eq:eq20}) are written in 
dimensionless units, where the temperature $T$
is in units of the critical temperature $T_c$; 
the order parameter $\psi$ is in units 
of $\psi_\infty(0)=\sqrt{\alpha(0)/\beta}$, 
where $\alpha$ and $\beta$ are two 
phenomenological constants; 
the distances are measured in units of the 
coherence length at zero temperature $\xi(0)$; 
the vector potential $\textbf{A}$ is in 
units of $H_{c2}(0) \xi(0)$, where $H_{c2}(0)$ 
is the upper critical field at zero 
temperature; the local magnetic field 
$\textbf{h}=\mbox{\boldmath $\nabla$}\times\textbf{A}$ 
is in units of $H_{c2}(0)$;
the current density is in units of $J_{GL} = \sigma\hbar/2e\xi(0)t_{GL}$ and we use $I_{GL} = J_{GL}\xi^2(0)$ as units of total current;
time is in units of the Ginzburg-Landau characteristic 
time $t_{GL}=\xi^2(0)/D$, where 
$D$ is the diffusion coefficient; the scalar potential 
is units of $\varphi_{GL}=H_{c2}(0)D/c$;
the material 
dependent parameter $\gamma=2\tau_E\Delta_0/\hbar$, 
where $\tau_E$ is the inelastic 
electron-collision time, and 
$\Delta_0$ is the gap in the Meissner state; 
$\kappa=\lambda(0)/\xi(0)$ is 
the Ginzburg-Landau parameter, where $\lambda(0)$ is 
the London penetration depth at zero temperature; 
the normal state electrical conductivity 
$\sigma$ is in units of $c^2/4\pi D\kappa^2$;
and finally, the constant $u$ 
is equal to 5.79, which is derived from first 
principles \cite{kramer1978theory}.

\begin{figure}[!t]
    \centering  
    \includegraphics[angle=0,width=\the\columnwidth]{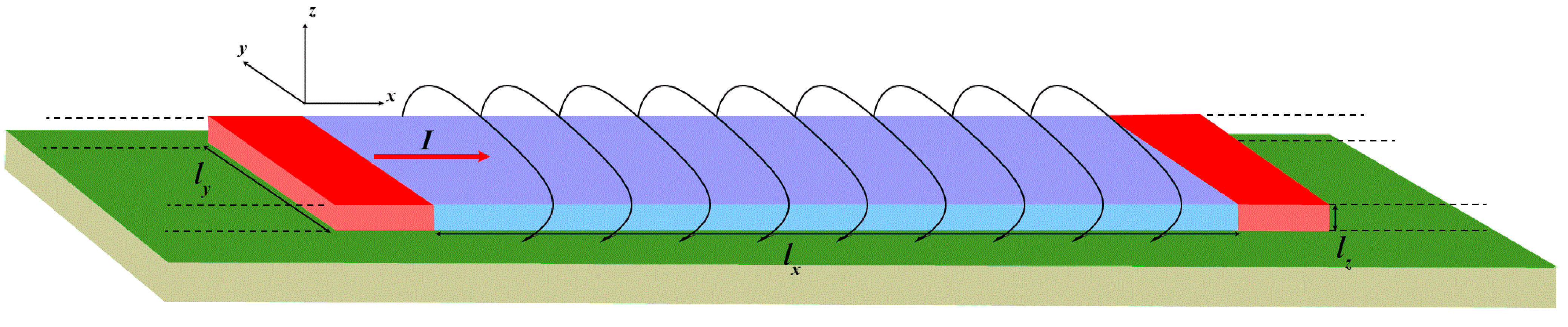}
    \caption{(Color online) Schematic view of a superconductor film (blue) 
    with dimensions specified in the figure. The red region 
    symbolizes a metallic film carrying a certain current 
    that is injected (ejected) into the superconductor through 
    two contacts, similar to a NSN junction. All these elements are on the top of a substrate (green).}
    \label{fig:fig_II_1}
\end{figure}

Additionally, to take into account the effects 
of the heat produced in the process of creation and 
annihilation of vortices, we couple the GTDGL 
equation with the heat diffusion equation \cite{vodolazov2005} 
given by:
\begin{equation}
    \mathcal{C}\frac{\partial T}{\partial t} = \zeta\nabla^2T +W
\label{eq:eq10}
\end{equation}
where $\mathcal{C}$ is the effective heat capacity, $\zeta$ is 
the effective heat conductivity coefficient. In this 
work, we have used $\mathcal{C} = 0.03$ and $\zeta = 0.06$. 
The term $W$ represents the power dissipated by the 
system. For gap superconductor it is given by \cite{duarte2017dynamics,duarte2023comparing,jing2018}:
\begin{equation}
    \resizebox{0.75\textwidth}{!}{$W = \sigma\textbf{E}^2+\frac{u}{\sqrt{1+\gamma^2|\psi|^2}}
        \left[ \left| \frac{\partial \psi}{\partial t}\right |^2
        +\frac{\gamma^2}{4}\left ( \frac{\partial |\psi|^2}{\partial t}\right )^2
        \right ]$},
        \label{eq:eq19}
\end{equation}
which is the generalization of the expression first determined 
for gapless superconductor in Ref.~\cite{schmid1966time}. Eqs.~\ref{eq:eq1}, \ref{eq:eq2}, \ref{eq:eq20}, and \ref{eq:eq10} must  solved to obtain the order parameter, the scalar and vector potentials and the local temperature.

Fig.~\ref{fig:fig_II_1} 
 illustrates the system under investigation, which consists of a 
 long film with a central cutout representing a 
 superconductor with dimensions ($l_x, l_y, l_z$) (blue region).
On both the right and left sides, two long metallic 
films (red regions) are present, where a fixed current density, 
denoted by $J_a$, is applied.
The total current injected (or ejected) at the metallic 
contacts is expressed by $I = l_y l_z J_a$. 
 Essentially, we have a NSN junction 
(normal metal - superconductor - normal metal), on top of a substrate (green region). 
To study solely the effects of temperature, we 
alter the system size according to the 
bath temperature $T_0$, the confinement being the 
same for all cases. For all temperatures, 
the dimensions are the same in units of $\xi(T)$, 
the values being $l_x = 12\xi(T)$, 
$l_y = 8\xi(T)$ and $l_z = 1.2\xi(T)$, where 
$\xi(T)=\xi(0)/\sqrt{1-T/T_c}$. For example, 
for $T_0 = 0.75 T_c$, we have  $l_x = 24\xi(0)$, 
$l_y = 16\xi(0)$ and $l_z = 2.4\xi(0)$.

The external current is introduced in the system through 
the boundary conditions for the scalar potential. At the 
superconductor/normal contact interfaces the scalar 
potential obeys $\hat{\textbf{n}}\cdot\mbox{\boldmath $\nabla$}\varphi = -J_a$, 
where $\hat{\textbf{n}}$ is the unit vector perpendicular 
to the interface and $J_a$ is the applied current density. 
At the remaining interfaces, we have 
$\hat{\textbf{n}}\cdot\mbox{\boldmath $\nabla$}\varphi = 0$.

To guarantee the superconducting current does not leave 
the stripe, the following condition must be imposed to 
the order parameter at the boundaries of the system:
\begin{equation}
    \hat{\textbf{n}}\cdot(-\mbox{i}\mbox{\boldmath $\nabla$}-\bm{A})\psi=\frac{\mbox{i}}{b}\psi\, ,
    \label{eq:eq17}
\end{equation}
where $b$ is the de Gennes extrapolation length; 
here $b$ is set to be equal to $20$ at the 
superconductor/normal contact interfaces and goes 
to infinity at the other interfaces.

At all interfaces of the $3D$ system, the local 
magnetic field $\textbf{h}$ is equal to the magnetic field $\textbf{H}$ 
produced by an uniform current density flowing in 
the superconducting film. This field is given by:
\begin{equation}
    \kappa^2\mbox{\boldmath $\nabla$}\times\textbf{H}=J_a\hat{\textbf{x}}\,.
    \label{eq:H_Ja}
\end{equation}
The solution of this equation can be found in the 
supplementary material of Ref.~\cite{cadorim2020ultra}.
Here, we must emphasize that, by using the solution of Eq.~\ref{eq:H_Ja} as our boundary conditions for the local field, we are disregarding demagnetization effects at the boundaries of the system. This approximation is a necessity  driven by the considerable computational demands associated with solving the Ampère's law in the surrounding regions of the superconducting system to obtain the stray fields. Although this is fully justified only for large values of $\kappa$, the qualitative features remain unchanged even for $\kappa \approx 1$.

Finally, we must specify the boundary conditions for 
the local temperature. 
At the interface superconductor/normal contact we 
set $T$ equals to the thermal bath temperature $T_0$. 
At the remaining interfaces, the boundary conditions 
are given by:
\begin{equation}
    T_{Out} = T_{In}-h_\alpha(T_{In}-T_0)
    \label{eq:eq18}
\end{equation}
which unifies both Dirichlet and Robin boundary conditions. 
Here, $T_{Out}$ is the temperature immediately 
outside the superconductor and $T_{In}$ is the temperature 
immediately inside the superconductor. For the lateral 
interfaces at $y = \pm l_y/2$ and the top surface at 
$z = l_z/2$ we have $h_\alpha = h_f$ whereas for the 
superconductor/substrate interface at $z = -l_z/2$, 
$h_\alpha = h_s$. The parameters $h_f$ and $h_s$ 
determine the strength of the heat removal through 
the surfaces, with $0$ meaning an insulator interface 
and $1$ an strong heat removal scenario. A graphical representation of how the values of $h_f$ and $h_s$ affects the local temperature distribution in the superconducting film is shown in the Supplementary Material.
In what follows, we numerically solve Eqs.~\ref{eq:eq1}, \ref{eq:eq2}, and \ref{eq:eq10} for the evolution of the order parameter, vector potential and local temperature, respectively. At each instant of time, Eq.~\ref{eq:eq20} is solved to obtain the scalar potential.

\section{\label{sec:sec3}Results and discussion}

The equations presented in Section \ref{sec:sec2} are 
discretized using the standard link-variable method, 
a technique detailed in Ref.~\cite{gropp1996numerical} (for further details see Ref.~\cite{cadorim2019}). 
Next, this algorithm is implemented using the 
Fortran 90 programming language and executed on a 
GPU (Graphics Processing Unit) with acceleration, 
employing a forward-time-central-space scheme. 
The grid spacing used is $\Delta x=\Delta y=\Delta z=0.4\xi(0)$, 
for $T_0=0.75T_c$, and $\Delta x=\Delta y=\Delta z=0.5\xi(0)$, 
for $T_0=0.84T_c$. Finally, the range $10\le \gamma \le 20$ 
proved suitable for most metals, such as 
Nb \cite{berdiyorov2009kinematic,kramer1978theory,watts1981nonequilibrium}; 
we adopted $\gamma=10$.

Before going further to the analysis of the results, let us elucidate 
how we derived the characteristic curves. The initial step 
is to compute the voltage, which we do as follows. Assuming 
that we measure the voltage between electrodes situated at 
$x=\pm l_x/2$ covering the width of the film, we determine the 
voltage along the $x$ direction through the following expression:
\begin{equation}
    U(t) = -\frac{1}{n_y-1}\sum_{j=2}^{n_y}\sum_{i=2}^{n_x}\,E_{x,i,j,n_z/2+1} \Delta x,
    \label{eq:eq21}
\end{equation}
where $n_x = l_x/\Delta x$, $n_y=l_y/\Delta y$, and $n_z = l_z/\Delta z$ 
are the number of unit cells of the $3D$ meshgrid in the 
$(x,y,z)$ directions, respectively.  
Then, we calculate the characteristic current-voltage by 
taking a time average of $U(t)$, expressed as:
\begin{equation}
    V = \frac{1}{\cal T}\,\int_0^{\cal T}\,U(t)\,dt, \label{eq:eq22}
\end{equation}
where ${\cal T}$  corresponds to the time required for an appropriate 
number of oscillations of $U(t)$.

Another quantity which is useful in the analysis of the results 
is the rate of heat transfer to the thermal bath. For this, we use 
the Fourier law:
\begin{equation}
    \frac{dq}{dt} = -h_{\alpha}\oint \mbox{\boldmath $\nabla$}T\cdot d\textbf{a}\,.
\end{equation}
Analogous to the voltage, this quantity oscillates throughout 
the resistive state. Then, we evaluate the time average 
of $\dot{q}$ as:
\begin{equation}
    \dot{Q} = \frac{1}{\cal T}\,\int_0^{\cal T}\,\dot{q}(t)\,dt, \label{eq:eq23}\,.
\end{equation}

We employ four distinct system dimensions according to 
the temperature as follows: 
(a) for $T_0 = 0.75T_c$, we set $l_x = 24\xi(0)$ 
and $l_y = 16\xi(0)$, exploring two different 
thicknesses, namely $l_z = 2.4\xi(0)$ 
and $l_z = 4.8\xi(0)$; (b) at $T_0 = 0.84T_c$, we adjust 
the dimensions to $l_x = 30\xi(0)$ and $l_y = 20\xi(0)$, 
again considering two distinct thicknesses, 
specifically $l_z = 3\xi(0)$ and $l_z = 6\xi(0)$.

    \subsection{Importance of the Heat Diffusion Equation in the 
                Study of the Resistive State}

    We begin the discussion of the results by highlighting the significance of incorporating the heat diffusion equation into the simulation of the resistive state in superconducting films.
    
    \begin{figure}[!h]
        \centering
        \includegraphics[angle=-90,width=\textwidth]{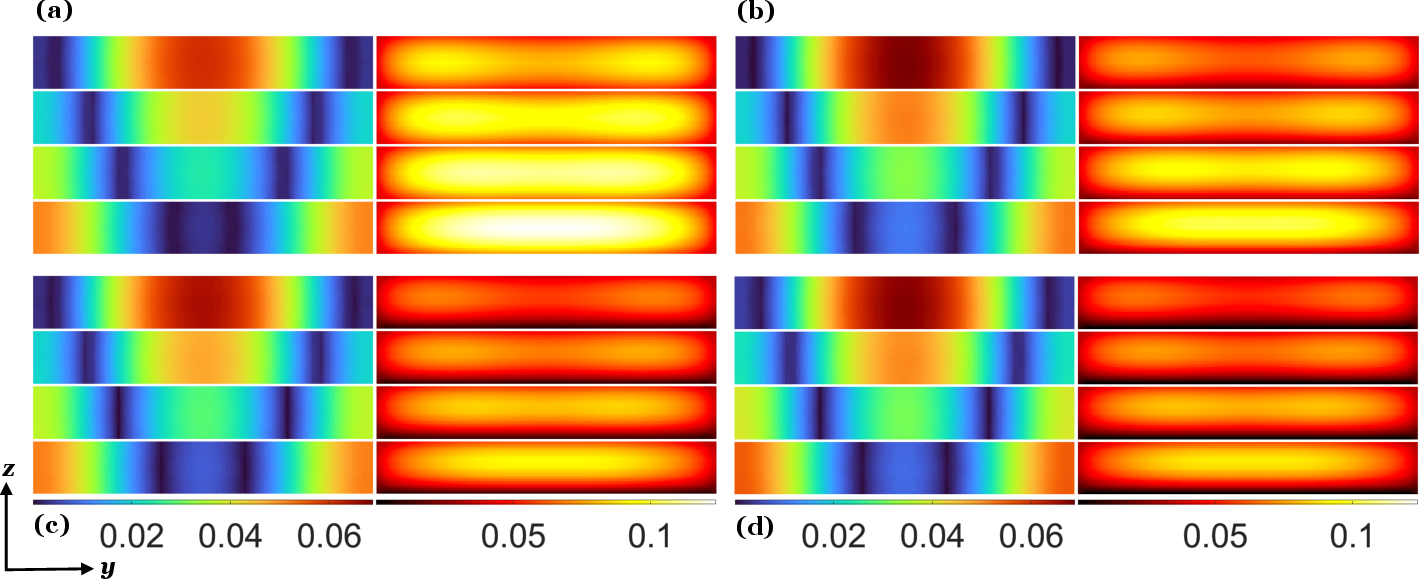}
        \caption{(Color online) Each panel exhibits colormaps of the intensity of order parameter $\psi$ (left column), and the increase of temperature $\Delta=T-T_0$ (right) in the $x=0$ plane ($yz$ plane): (a) $h_f = 0.25,\,h_s = 0.25$; (b) $h_f = 0.25,\,h_s = 0.50$; (a) $h_f = 0.25,\,h_s = 0.75$; (a) $h_f = 0.25,\,h_s = 1.00$; for all panels $J_a = 0.049J_{GL}$, $T_0 = 0.75T_c$, $l_x = 24\xi(0)$, $l_y = 16\xi(0)$, $l_z = 2.4\xi(0)$. For the order parameter, deep blue indicates regions where the order parameter is significantly degraded, implying a substantial suppression of superconductivity. This phenomenon occurs due to the presence of vortex-antivortex pairs penetrating the superconductor, as indicated by the vertical blue stripes. On the other hand, deep red represents areas where superconductivity is less suppressed, suggesting lesser degradation of the order parameter. For the temperature, red corresponds to regions of lower temperature in the superconductor, where as white  denotes areas where the temperature is higher, usually associated with the movement of v-av pairs from their nucleation to their annihilation at the center of the superconductor.}
        \label{fig:fig15}
    \end{figure}

    In Fig.~\ref{fig:fig15} we present the intensity 
    of the order parameter 
    and the temperature for $h_f=0.25$ and four values of the substrate 
    $h_s$. Each panel shows that as the strength of the substrate increases 
    (from panel (a) to (d)), there is a tendency for the temperature of 
    the superconductor to decrease. This demonstrates the effectiveness 
    of a robust substrate in dissipating the heat generated by the 
    movement of v-av pairs. The movement of these pairs is a 
    significant source of heat due to the resistance they offer 
    to the electrical current.
    
    The results from the simulations indicate that the 
    choice and implementation of an efficient substrate are key for 
    thermal control in superconductors, particularly under high DC 
    current conditions.  

    \begin{figure}[!h]
        \centering
        \includegraphics[angle=-90,width=0.75\textwidth]{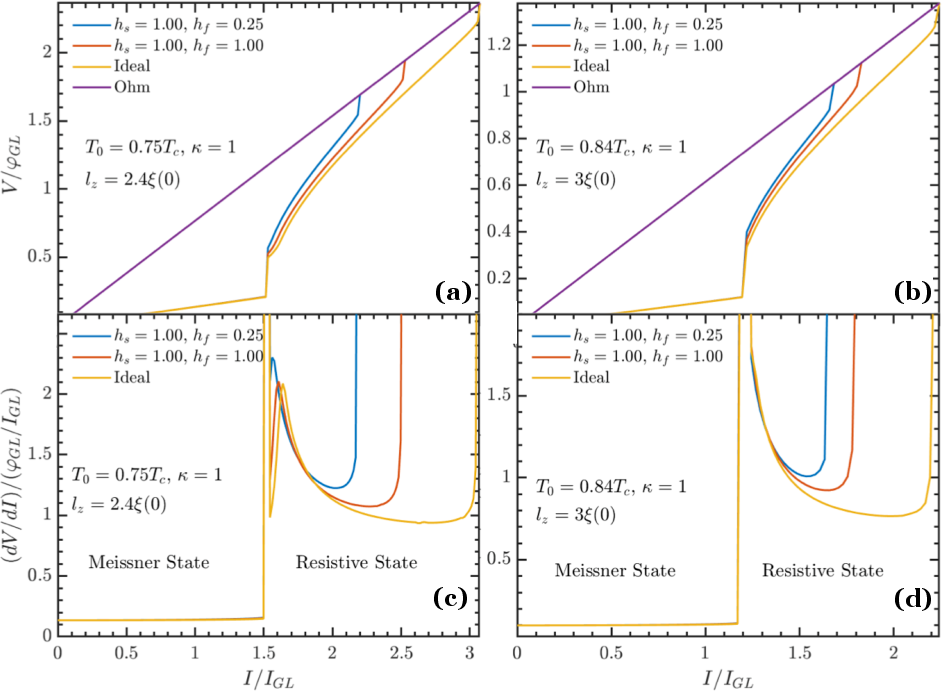}
        \caption{(Color online) \textit{IV} and \textit{IR} characteristic curves for $T_0 = 0.75 T_c$ (panels (a) and (c), respectively) and $T_0 = 0.84 T_c$ (panels (b) and (d), respectively). Yellow curve represents the results for the simulation which does not solve the heat diffusion equation, while red and blue curves represent results for simulations in which the heat equation is solved with parameters $h_s = 1.00$, $h_f = 1.00$ and $h_s = 1.00$, $h_f = 0.25$, respectively. The purple curve is the Ohm curve for the voltage at the normal state.}
        \label{fig:fig_III_A_1}
    \end{figure}

    In Fig.~\ref{fig:fig_III_A_1}, we present the characteristic curves for current-voltage (\textit{IV}) (panels (a) and (b)) and current-resistance (\textit{IR}) (panels (c) and (d)). These curves correspond to two distinct temperature values: $T_0 = 0.75 T_c$ (panels (a) and (c)) and $T_0 = 0.84 T_c$ (panels (b) and (d)). The assigned values for $l_z$ are given by $2.4 \xi(0)$ and $3 \xi(0)$ for $T_0 = 0.75 T_c$ and $T_0 = 0.84 T_c$, respectively.

    In this figure, the yellow curves illustrate the results when the heat diffusion equation remains unsolved, that is, the temperature is considered homogeneous throughout the superconductor. Hereafter, we refer to this case as the ideal one. In contrast, the remaining two curves incorporate heat dissipation in the simulations. 
    The red curve stands for $h_s = 1.00$ and $h_f = 1.00$, which from now on we will denominate by strong heat removal scenario, while the blue curve represents a scenario with a less efficient thermal bath, where $h_s = 1.00$ and $h_f = 0.25$.

    In all cases, the behavior of the system unfolds as follows. The superconductor initiates from the Meissner state, characterized by the applied current flowing through the stripe in the form of a dissipationless supercurrent. The finite voltage observed in Fig.~\ref{fig:fig_III_A_1} emerges due to the presence of the normal contacts. Upon reaching a critical current denoted as $I_{c1}$, the superconductor undergoes a transition to the resistive state. In this state, periodic nucleation of \textit{v-av} pairs occurs at the sample boundaries, followed by their annihilation at the center.
    As described in Ref.~\cite{cadorim2023}, due to the finite thickness of the film, a \textit{v-av} pair is composed by curved vortex and antivortex which form a closed vortex before annihilating themselves.
    This dynamics was described in Ref.~\cite{cadorim2023} with an animation presented in the Supplement Material of this reference.
    As the applied current reaches another critical value, $I_{c2}$, superconductivity is entirely suppressed, leading the superconductor to the normal state.

    As can be seen, the transition from the Meissner to the resistive state is unaffected by heat diffusion. The absence of vortices in the Meissner state implies a lack of substantial heat dispersion sources within the system during this phase, thus $I_{c1}$ is the same for all scenarios. In opposition, the value of $I_{c2}$ experiences significant modulation due to the involvement of the heat diffusion equation. Notably, its magnitude is significantly greater when heat considerations are omitted from the simulations. With the inclusion of heat generated by the \textit{v-av} pairs, the dynamics of the resistive state undergo a clear change. This modification manifests as a reduction in the duration of the resistive state, since the rise in local temperature becomes a contributing factor to the suppression of the superconducting state at lower currents, specially for lower values of $T_0$.   

    The impacts of heat diffusion also manifest in the resistance of the system. Just after $I_{c1}$, the resistance remains nearly unchanged regardless of whether the heat diffusion equation is taken into account or not. However, a distinct shift occurs when currents approach the vicinity of $I_{c2}$, where scenarios involving heat dissipation exhibit higher resistance values. The heat dissipation originating from \textit{v-av} pairs annihilation rises proportionally with the applied current, rendering the heat diffusion equation progressively more significant within these regions.
    Consequently, as heat depletes the superconducting state near the critical current $I_{c2}$, the overall medium becomes less conductive to the smooth flow of superconducting current. This accounts for the observed increase in resistance in these situations.

    \begin{figure}[!h]
        \centering
        \includegraphics[angle=90,width=0.75\textwidth]{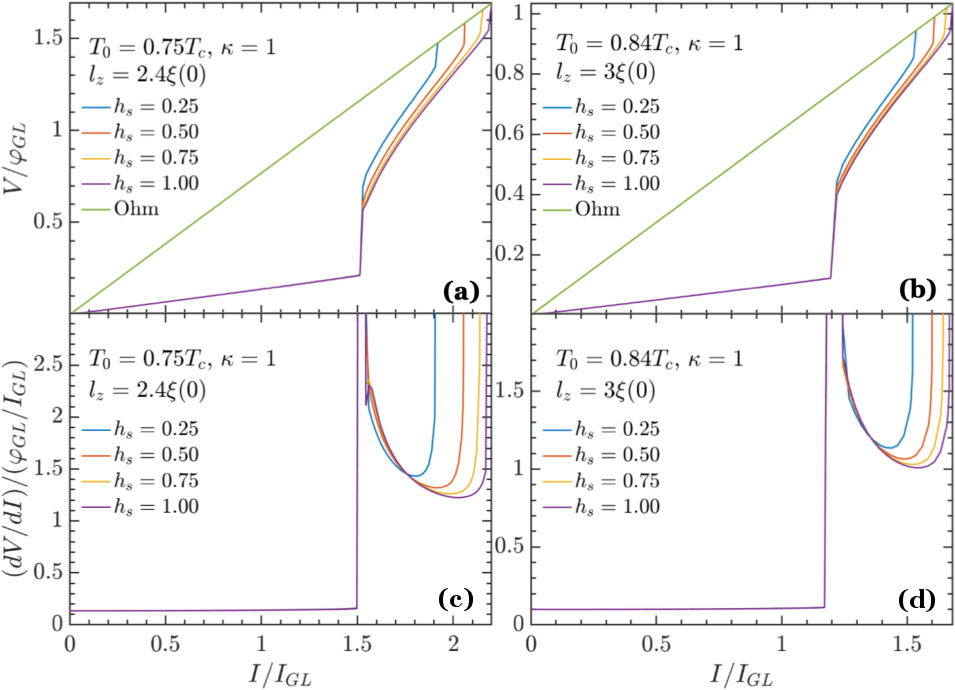}
        \caption{(Color online) \textit{IV} and \textit{IR} curves for $T_0 = 0.75 T_c$ (panels (a) and (c), respectively) and $T_0 = 0.84 T_c$ (panels (b) and (d), respectively). Each curve represents the results for a different value of $h_s$, as specified in the legends. Green curve is the Ohm curve for the voltage at the normal state.}
        \label{fig:fig_III_B_1}
    \end{figure}

    \subsection{Efficiency of the Substrate}
    
    Having established the importance of accounting for heat diffusion to accurately describe the resistive state, we now examine how the heat removal process influences the behavior of the system. For this purpose, we varied the parameter $h_s$, maintaining $h_f = 0.25$ fixed. In other words, we change the efficiency of heat removal from the superconductor from weak to strong.

    In Fig.~\ref{fig:fig_III_B_1}, we present the \textit{IV} and \textit{IR} curves for two distinct temperatures ($T_0 = 0.75 T_c$ and $T_0 = 0.84 T_c$) and for four different values of $h_s$. As described in the preceding Section, the heat produced by the \textit{v-av} pairs plays a significant role in the reduction of superconductivity, and thus increasing the resistance. This phenomenon is illustrated in panels $(c)$ and $(d)$ of Fig.~\ref{fig:fig_III_B_1}, where the resistance becomes lower near the critical current $I_{c2}$ as the efficiency of heat removal from the sample improves.

    \begin{figure}[!h]
        \centering
        \includegraphics[width=0.75\textwidth]{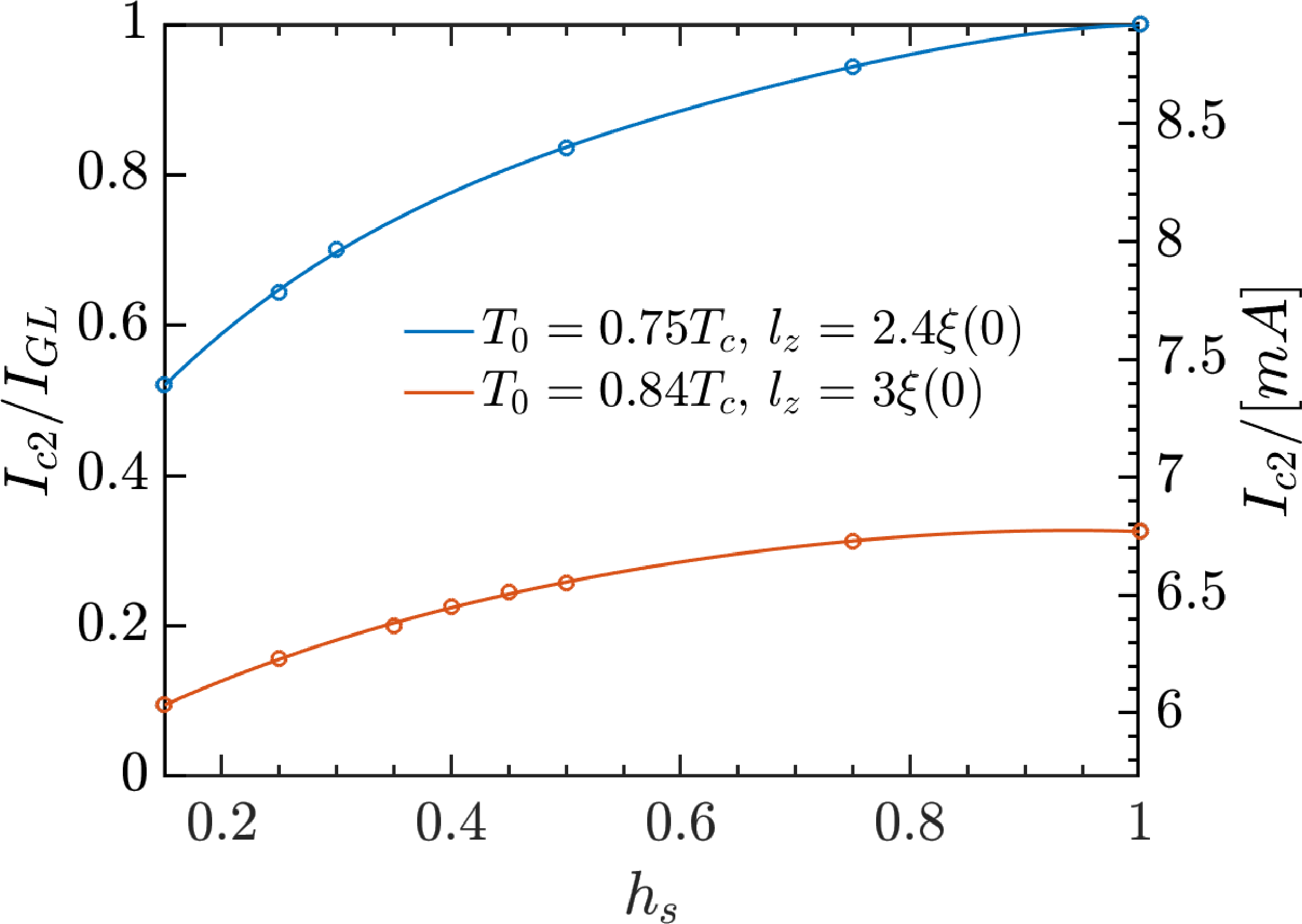}
        \caption{(Color online) Critical current $I_{c2}$ as a function of $h_s$ for $T_0 = 0.75 T_c$ (blue curve) and $T_0 = 0.84 T_c$ (red curve). Left $y$ axis shows the critical current in dimensionless units, while the right $y$ axis shows the current in real units, using Nb parameters $\xi(0) = 39nm$ and $\lambda(0) = 50nm$ as a reference.}
        \label{fig:fig_III_B_4}
    \end{figure}

    As anticipated, the efficiency of heat removal also directly impacts the critical current for the transition to the normal state. To facilitate the visualization of the dependence of $I_{c2}$ on the heat removal, Fig.~\ref{fig:fig_III_B_4} shows $I_{c2}$ as a function of $h_s$ for both temperature values. As can be seen, the critical current exhibits a monotonic rise with increasing $h_s$, although this variation is not linear. Instead, the change in $I_{c2}$ becomes particularly pronounced for smaller values of $h_s$, whereas the critical current presents a gradual increase as we approach a scenario characterized by strong heat removal conditions. To give an idea of what this critical current variation means in a real system, the right $y$ axis presents $I_{c2}$ in real units, 
    with Nb parameters $\xi(0) = 39$ nm and $\lambda(0) = 50$ nm serving as a reference    \cite{poole2014superconductivity}. The value of $\kappa$ for this material is close to the one used in our simulations. 
    However, we note that, since the parameters used in the heat diffusion equation are general and may not representative of a Nb film. As  
    such values should be considered solely as an estimate of how $I_{c2}$ depends on the substrate efficiency.    

    By comparing the two curves, a clear distinction emerges, that is, the change in $I_{c2}$ is more pronounced for $T_0 = 0.75T_c$. This difference can be attributed to the greater heat dissipation originating from the \textit{v-av} pairs in this particular scenario. Consequently, the efficiency of the heat removal process becomes more important for the evolution of the superconducting state.
    
\begin{figure}[!h]
        \centering
        \includegraphics[angle=-90,width=0.75\textwidth]{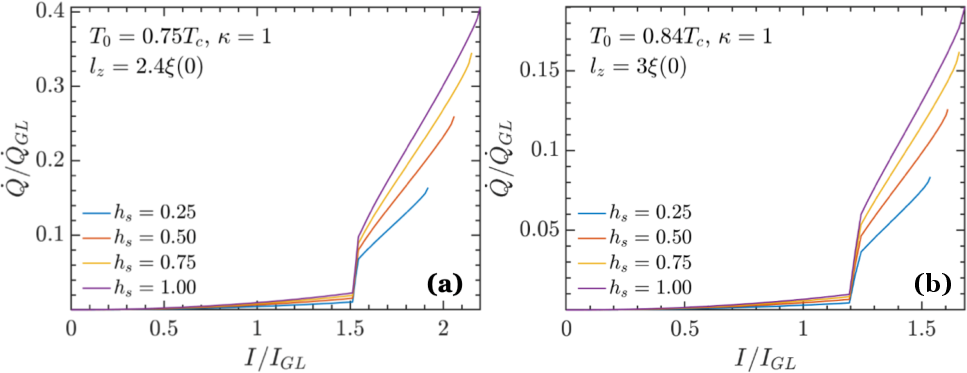}
        \caption{(Color online) Rate of heat transfer as a function of the applied current for $T_0 = 0.75 T_c$ (panel (a)) and $T_0 = 0.84 T_c$ (panel (b)). Each curve represents the results for a different value of $h_s$, as specified in the legends.}
        \label{fig:fig_III_B_3}
    \end{figure}

    To show that the dissipated heat indeed increases with decreasing system temperature, Fig.\ref{fig:fig_III_B_3} provides a depiction of the heat transfer rate for all cases illustrated in Fig.\ref{fig:fig_III_B_1}. As it can be seen, at $T_0 = 0.75 T_c$, the heat extracted from the superconductor surpasses that at $T_0 = 0.84 T_c$ by a factor of more than two. This difference can be attributed to the nature of the superconducting medium, which becomes increasingly resistant to the passage of a vortex as the superconductivity grows more robust at $T_0 = 0.75 T_c$, thus leading to a larger dissipated heat.
    Furthermore, Fig.~\ref{fig:fig_III_B_3} shows that the heat transfer increases with $h_s$, as expected from the previous analysis.

    \begin{figure}[!h]
        \centering
        \includegraphics[angle=-90,width=0.75\textwidth]{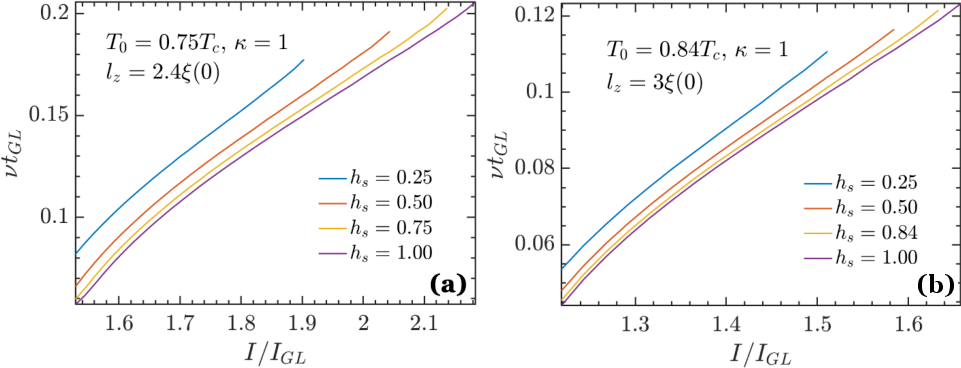}
        \caption{(Color online) Average frequency of the process of creation and annihilation of \textit{v-av} pairs as a function of the applied current. Panel $(a)$ shows the result for $T_0 = 0.75 T_c$ and panel $(b)$ for $T_0 = 0.84 T_c$. Each curve represents a different $h_s$, as described in the legends.}
        \label{fig:fig_III_B_2}
    \end{figure}

    To explore the influence of heat removal efficiency on the process of \textit{v-av} pair creation and annihilation, Fig.~\ref{fig:fig_III_B_2} presents the frequency of this phenomenon as a function of the applied current. This is demonstrated for different values of $h_s$ and two distinct temperatures, namely, $T_0 = 0.75 T_c$ and $T_0 = 0.84 T_c$. It becomes clear that, for a given temperature, the frequency of the vortex pair dynamics rises as heat removal from the superconductor becomes less effective. This can be attributed to the decreased viscosity of suppressed superconductivity regions in regard to the motion of vortices. For lower $h_s$, a significant proportion of the dissipated heat remains within the superconductor. This heat accumulation increases the local temperature, amplifying the velocity of the \textit{v-av} pairs.
    In addition, the authors of Ref.~\cite{vodolazov2007}
    argue that the relaxation time of the order parameter
    depends on the temperature as $\tau_{|\psi|} = 1/(T_c-T_0)$.
    Indeed, as depicted in Fig.~\ref{fig:fig_III_B_2}, the frequency of v-av pair collisions rises with decreasing $T_0$.
    Consequently, resistance is expected to be greater at
    lower temperatures. This tendency is clearly observed in
    panels $(c)$ and $(d)$ of Fig. ~\ref{fig:fig_III_A_1}, even in the ideal scenario. This suggests that the observed effect cannot be solely attributed to heat diffusion.

    \subsection{Effect of the Film Thickness}

\begin{figure}[!h]
        \centering
        \includegraphics[angle=90,width=0.75\textwidth]{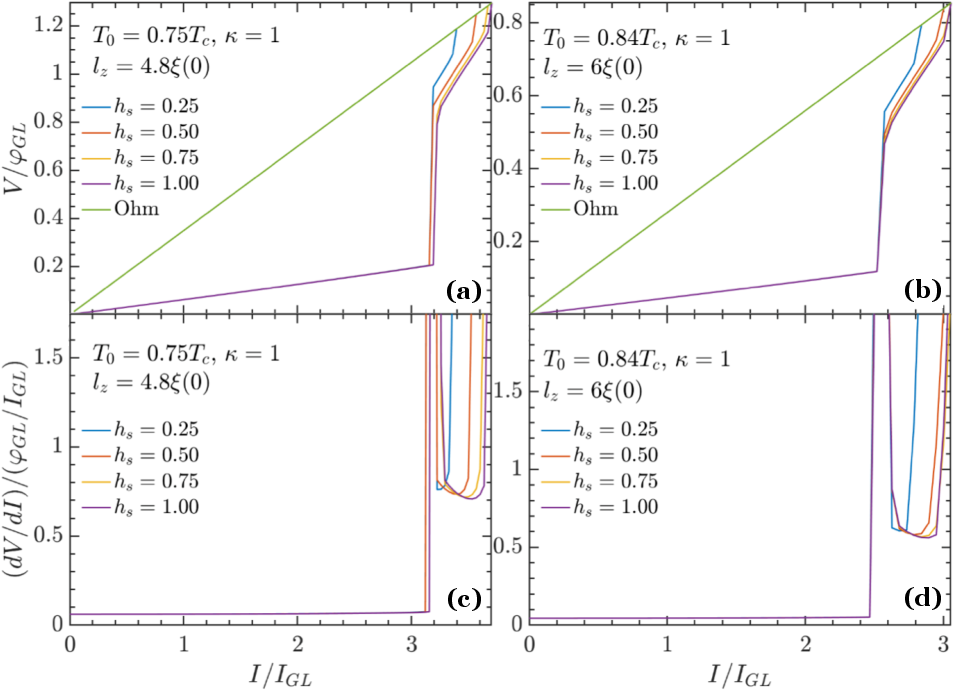}
        \caption{(Color online) \textit{IV} and \textit{IR} curves for $T_0 = 0.75 T_c$, $l_z = 4.8 \xi(0)$ (panels (a) and (c), respectively) and $T_0 = 0.84 T_c$, $l_z = 6.0 \xi(0)$ (panels (b) and (d), respectively). Each curve represents the results for a different value of $h_s$, as specified in the legends. Green curve is the Ohm curve for the voltage at the normal state.}
        \label{fig:fig_III_C_1}
    \end{figure}

\begin{figure}[!h]
        \centering
        \includegraphics[angle=-90,width=0.75\textwidth]{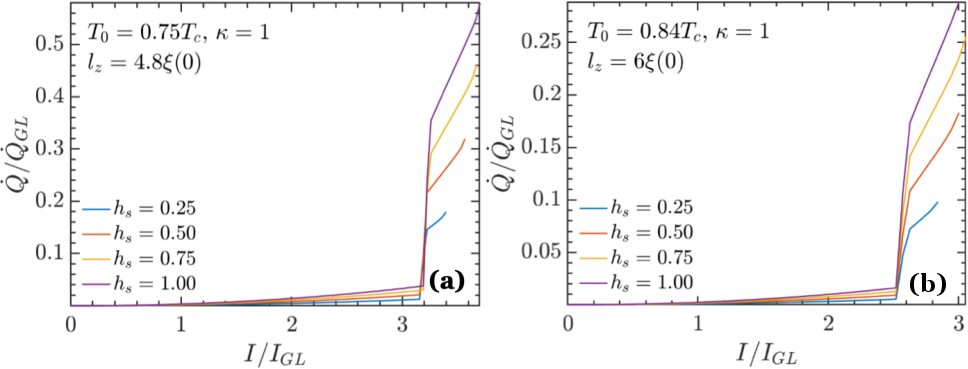}
        \caption{(Color online) Rate of heat transfer as a function of the applied current for $T_0 = 0.75 T_c$, $l_z = 4.8 \xi(0)$ (panel (a)) and $T_0 = 0.84 T_c$, $l_z = 6.0 \xi(0)$ (panel (b)). Each curve represents the results for a different value of $h_s$, as specified in the legends.}
        \label{fig:fig_III_C_3}
    \end{figure}

\begin{figure}[!h]
        \centering
        \includegraphics[angle=-90,width=0.75\textwidth]{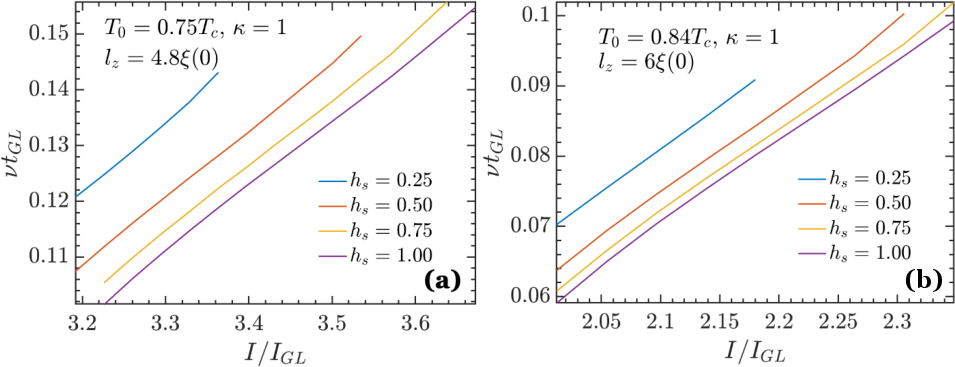}
        \caption{(Color online) Average frequency of the process of creation and annihilation of \textit{v-av} pairs as a function of the applied current. Panel $(a)$ shows the result for $T_0 = 0.75 T_c$, $l_z = 4.8 \xi(0)$ and panel $(b)$ for $T_0 = 0.84 T_c$, $l_z = 6.0 \xi(0)$. Each curve represents a different $h_s$, as described in the legends.}
        \label{fig:fig_III_C_2}
    \end{figure}

    We now investigate how the thickness of the film influences the characteristics of the resistive state. While keeping the previously defined $l_x$ and $l_y$ values, we adjust $l_z$ to be $4.8 \xi(0)$ for $T_0 = 0.75 T_c$ and $6 \xi(0)$ for $T_0 = 0.84 T_c$. This is depicted in Figs.~\ref{fig:fig_III_C_1}, \ref{fig:fig_III_C_3}, and \ref{fig:fig_III_C_2}, which respectively show the \textit{IV} and \textit{IR} curves, the heat transfer rate, and the frequency of the creation and annihilation process of \textit{v-av} pairs as functions of the applied current.

    These figures show that the system characteristics described in the preceding Section remain unchanged, despite doubling the thickness of the superconductor. However, the figures reveal a new feature, namely that the parameter $h_s$ is more important for the quantitative properties of the system. This observation is particularly pronounced for smaller values of $h_s$,
    where heat is less effective and the dissipated heat becomes more important to the evolution of the superconducting state.

    This effect stems from two factors. Firstly, by comparing Figs.~\ref{fig:fig_III_B_3} and \ref{fig:fig_III_C_3}, we can see that the heat dissipated by the vortex motion is larger for thicker films. Once the superconducting medium is viscous to the vortex flux flow, a larger thickness yields a larger rate of heat transfer. Secondly, a larger thickness means it is more difficult for the heat to be removed from the superconductor by the top and bottom surfaces, thus increasing the importance of the value of $h_s$. In addition, the extension of the resistive state, relatively to the Meissner state, is much less than for the thinner superconducting films.

    \begin{figure}[!h]
        \centering
        \includegraphics[angle=90,width=0.75\textwidth]{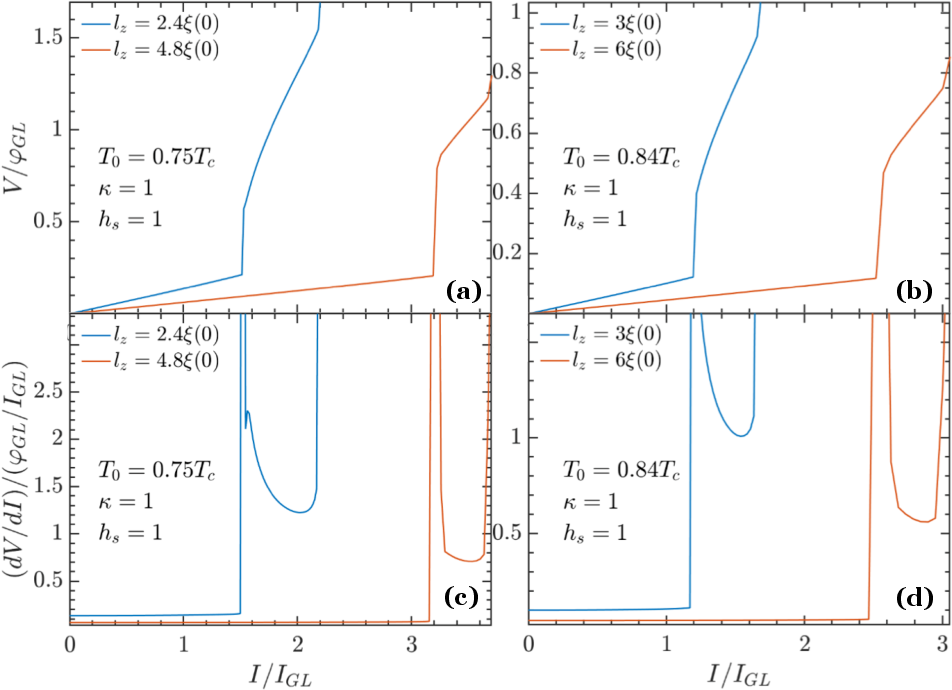}
        \caption{(Color online) \textit{IV} and \textit{IR} curves for $T_0 = 0.75 T_c$ with $l_z = 2.4 \xi(0)$ and $4.8 \xi(0)$ (panels (a) and (c), respectively) and $T_0 = 0.84 T_c$ with $l_z = 3.0 \xi(0)$ and $6.0 \xi(0)$ (panels (b) and (d), respectively). In all cases, we have set $h_s = 1.00$.}
        \label{fig:fig_III_C_4}
    \end{figure}

    To facilitate a detailed comparison between the different thicknesses, Fig.~\ref{fig:fig_III_C_4} presents the \textit{IV} and \textit{IR} curves for $h_s = 1.00$ of the two simulated thickness values for each temperature.
    The disparity between $I_{c1}$ and $I_{c2}$ for different thicknesses occurs because the transitions to the resistive and to the normal state are governed by the applied current density $J_a$, and their critical values are approximately the same for both films.
    The second feature that comes apparent in this figure is the difference in the resistance between the thicknesses, with the thicker films displaying a smaller resistance. Although thicker films dissipate more heat, they also carry a larger amount of current, which gives origin to a smaller resistance.

    \subsection{Hysteresis Loops}

    After investigating the behavior of the resistive state for various heat removal scenarios and different superconductor thicknesses under increasing applied currents, we now turn our attention to its response as the current is gradually reduced.

    \begin{figure}[!h]
        \centering
        \includegraphics[angle=270,width=0.75\textwidth]{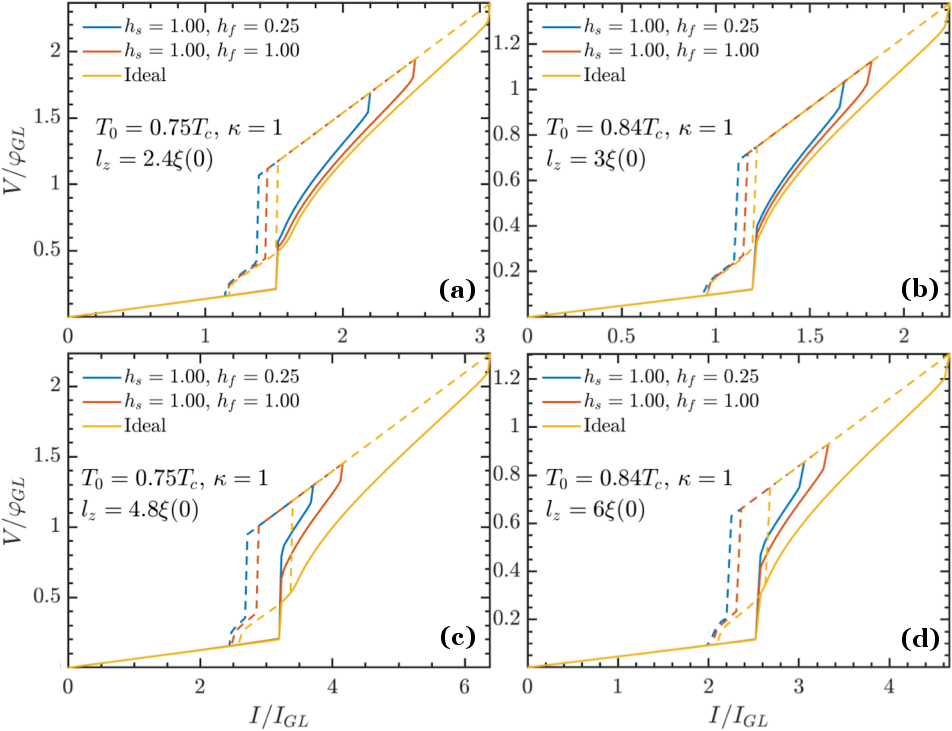}
        \caption{(Color online) \textit{IV} curves for the upward (solid lines) and downward (dashed lines) sweep directions of the applied current. Panel $(a)$ corresponds to $T_0 = 0.75 T_c$, $l_z = 2.4 \xi(0)$; panel $(b)$ to $T_0 = 0.84 T_c$, $l_z = 3.0 \xi(0)$; panel $(c)$ to $T_0 = 0.75 T_c$, $l_z = 4.8 \xi(0)$; panel $(d)$ to $T_0 = 0.84 T_c$, $l_z = 6.0 \xi(0)$. Each panel shows the \textit{IV} curve for the case where the heat diffusion equation is not solved (yellow line), for an strong heat removal scenario with $h_s = 1.00$, $h_f = 1.00$ (red line) and for $h_s = 1.00$ and $h_f = 0.25.$ (blue line).}
        \label{fig:fig_III_D_1}
    \end{figure}

    In Fig.~\ref{fig:fig_III_D_1}, the \textit{IV} curves for three distinct systems are displayed. For each system, computed voltage values are presented for both directions of current sweep, encompassing three distinct heat removal scenarios. The first scenario involves the absence of numerical solution for the heat diffusion equation (yellow curve), while the subsequent two scenarios incorporate heat considerations. These include the strong scenario with $h_f = 1.00$ and $h_s = 1.00$ (red curve), as well as the scenario with $h_f = 0.25$ and $h_s = 1.00$ (blue curve).

    Panels $(a)$ and $(b)$ of Fig.~\ref{fig:fig_III_D_1} illustrate the hysteresis loops for two distinct temperatures: $T_0 = 0.75 T_c$, with $l_z = 2.4 \xi(0)$, and $T_0 = 0.84 T_c$, with $l_z = 3 \xi(0)$. Notably, as observed from the yellow curves, in the absence of the heat diffusion equation, the system goes from the normal state back to the resistive state at a critical current $I_{c2}^*$ that closely aligns with $I_{c1}$. Below this critical current, the system remains in the resistive state until it reaches $I_{c1}^*$, marking the point where it reverts to the Meissner state.
    
    In contrast, for the red and blue curves where the heat diffusion equation is resolved, we can see that the value of $I_{c2}^*$ is lower. This discrepancy arises due to the interaction between the recovering superconducting state and the heat generated by the moving vortices. This heat avoids the stabilization of the superconducting state, consequently leading to a reduction in the critical current necessary for the onset of the resistive state.

    Furthermore, it is worth noting that the distinction in $I_{c2}^*$ between the two cases considering heat and the case where the heat diffusion equation is neglected increases with the thickness of the superconductor.
    The difference between $I_{c2}^*$ of the two cases where heat is taken into account and of the case where the heat diffusion equation is not solved increases with the thickness of the superconductor (compare panels $(a)$ and $(b)$ with panels $(c)$ and $(d)$, respectively).
    As discussed in the last Section these results reinforce the fact that the diffusion equation becomes more important for thicker films, once the dissipated heat is larger while heat removal also becomes less efficient.
    
    On the other hand, it is noticeable that $I_{c1}^*$ is also diminished for the two systems where the heat diffusion equation is accounted for. In these scenarios, the heat generated by the vortices serves to reduce the energy barrier that the \textit{v-av} pairs must overcome to penetrate the superconductor. This implies the presence of the resistive state at lower current values. Similarly, the system characterized by $h_f = 0.25$ exhibits a smaller $I_{c1}^*$ compared to the system with $h_f = 1.00$, as the latter benefits from a more efficient heat removal mechanism.
    
    \subsection{Effect of the Ginzburg-Landau Parameter}

    \begin{figure}[!h]
        \centering
        \includegraphics[angle=90,width=0.75\textwidth]{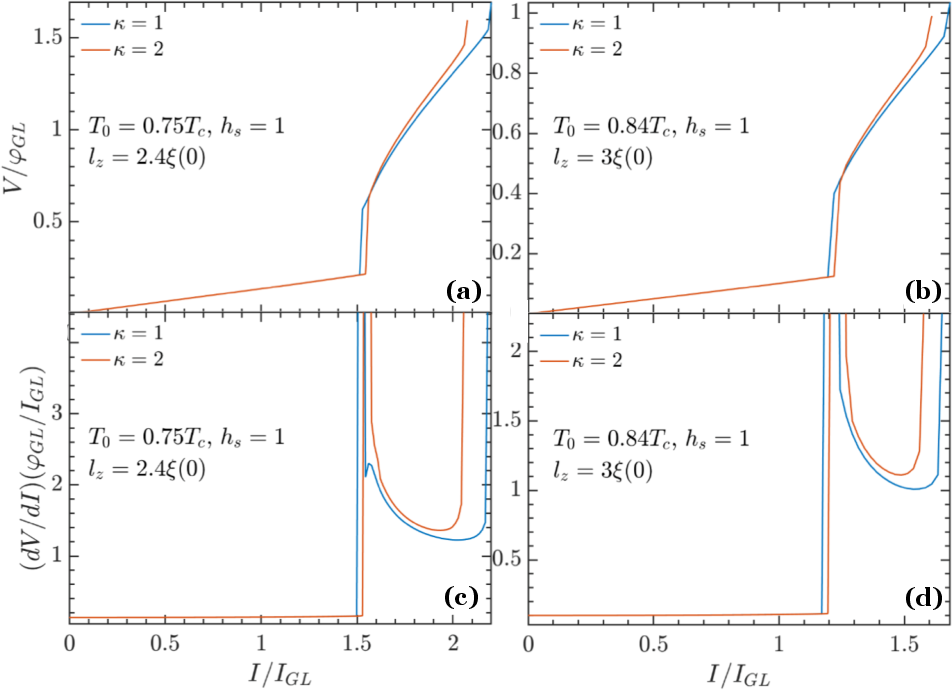}
        \caption{(Color online) \textit{IV} and \textit{IR} curves for $\kappa = 1.0$ and $2.0$. Panels $(a)$ and $(c)$ corresponds to $T_0 = 0.75 T_c$ and $l_z = 2.4 \xi(0)$, while panels $(b)$ and $(d)$ corresponds to $T_0 = 0.84 T_c$ and $l_z = 3.0 \xi(0)$. In all cases, we have set $h_s = 1.00$.}
        \label{fig:fig_III_E_1}
    \end{figure}

    In Fig.~\ref{fig:fig_III_E_1}, a comparison between the \textit{IV} and \textit{IR} curves is presented for two different values of the Ginzburg-Landau parameter, $\kappa = 1$ and $\kappa = 2$. Despite the similarity in the critical current $I_{c1}$ that marks the beginning of the resistive state for both $\kappa$ values, we can see that $I_{c2}$ decreases and the resistance increases as we increase $\kappa$.

\begin{figure}[!t]
        \centering
        \includegraphics[angle=270,width=0.75\textwidth]{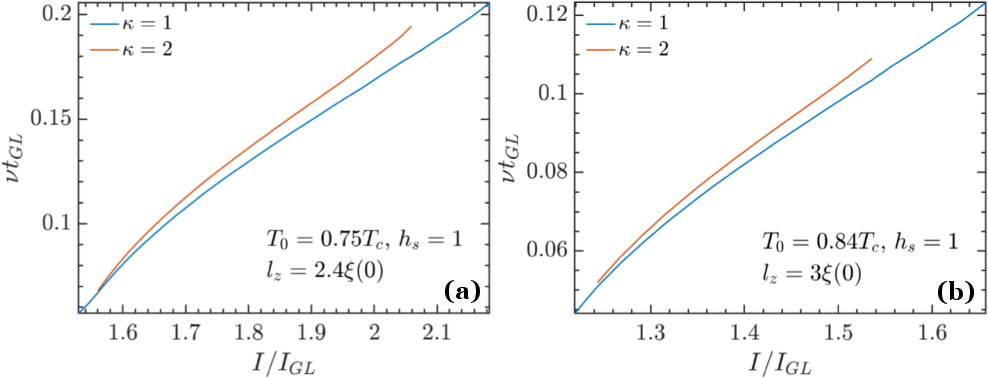}
        \caption{(Color online) Average frequency of the process of creation and annihilation of \textit{v-av} pairs as a function of the applied current. Panel $(a)$ shows the result for $T_0 = 0.75 T_c$, $l_z = 2.4 \xi(0)$ and panel $(b)$ for $T_0 = 0.84 T_c$, $l_z = 3.0 \xi(0)$. In each panel, the blue and the red curves represent $\kappa = 1.0$ and $\kappa = 2.0$, respectively.}
        \label{fig:fig_III_E_2}
    \end{figure}

    As depicted in Fig.\ref{fig:fig_III_E_2}, a larger $\kappa$ value corresponds to an increased frequency in the process of creating and annihilating \textit{v-av} pairs. Once a higher frequency also means a larger amount of dissipated heat, we can see that this feature is the responsible for the larger resistance and smaller $I_{c2}$ displayed in Fig.\ref{fig:fig_III_E_1}. The relationship between these quantities is further confirmed by noticing that when the frequency for the two values of $\kappa$ are approximately the same for applied currents near $I_{c1}$, the calculated voltage is also almost equal for both cases. On the other hand, as the applied current approaches $I_{c2}$, the frequency for $\kappa = 2$ increases at a larger rate than its counterpart for $\kappa = 1$, which is followed by a larger voltage for the former case.

\subsection{Comparison with the $2D$ model}

Finally, let us investigate the importance of the three-dimensional nature of the method developed here. In other words, we will study how the predictions of our model differ from the ones obtained by the conventional two-dimensional model, very often applied in the literature. In the $2D$ case, Eqs.~\ref{eq:eq1}, \ref{eq:eq2} and \ref{eq:eq20} are solved in their two-dimensional form, while Eq.~\ref{eq:eq10} is replaced by \cite{vodolazov2005}:

\begin{equation}
    \mathcal{C}\frac{\partial T}{\partial t} = \zeta\nabla^2T-\eta(T-T_0)+W ,
\label{eq:eq24}
\end{equation}
here, the second term, absent in our $3D$ model, mimics the heat removal from the superconductor by the substrate. As boundary conditions, the temperature is set to be equal to the bath temperature $T_0$ in all edges. As defined in Ref.~\cite{vodolazov2005}, we set $\eta = 0.002$, which corresponds to a strong heat removal.

\begin{figure}[!h]
        \centering
        \includegraphics[width=0.75\textwidth]{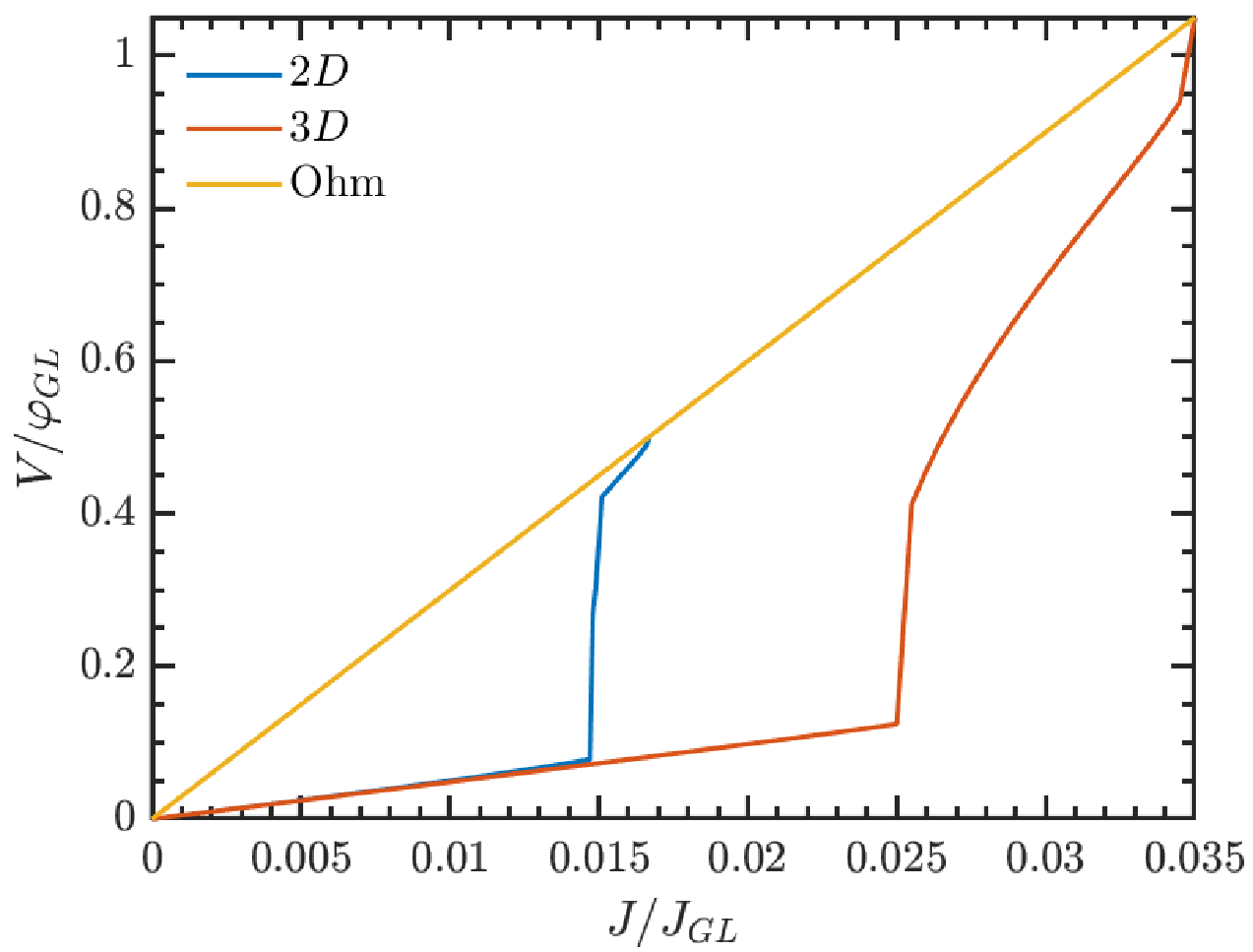}
        \caption{(Color online) Comparison of the \textit{IV} characteristics obtained with the $2D$ (blue curve) and $3D$ (red curve) models. In the $3D$ model, we have $T= 0.84 T_c$, $l_z = 1.0\xi(0)$ and $h_s = h_f = 1.00$, while in the $2D$ model, we have $\eta = 0.002$.}
        \label{fig:fig_III_F_1}
    \end{figure}

In Fig.~\ref{fig:fig_III_F_1}, we show the \textit{IV} curves for the $2D$ (blue curve) and $3D$ (red curve) models. The thickness of the superconducting film is $l_z = 1.0\xi(0)$ and the bath temperature $T = 0.84 T_c$.
As one can see, the quantitative difference between the results given by each method is clear. Both critical currents are significantly altered, as the resistive state begins at a much smaller current in the $2D$ case and $J_{c2}$ varies even more, giving rise to a small resistive state range of currents.

It is important to note the thickness of the $3D$ film is smaller than the coherence length of the system at the bath temperature, which is usually the condition used to justify the use of $2D$ equations. As shown in Fig.~\ref{fig:fig_III_F_1}, though, while the $2D$ model remains valid to investigate the qualitative behavior of the vortex dynamics in the resistive state of superconducting films, its quantitative predictions fall short when compared to our more complete $3D$ model. This has important consequences to the design of novel superconducting devices, where the precise knowledge of the critical properties of the system under investigation is needed, thus making it necessary the utilization of the $3D$ model.

\section{\label{sec:sec4}Concluding Remarks}

In conclusion, this study highlights the importance 
of considering the effects of heat diffusion, heat 
removal efficiency, film thickness, and the Ginzburg-Landau 
parameter when analyzing the resistive state in superconducting 
films. These findings contribute to a better understanding of the 
behavior of superconducting materials under a variety of  
conditions and offer insights that could have practical 
implications for the design and applications of superconducting 
devices, especially regarding the critical parameters. As we have shown, the three-dimensional model used here is of great importance to the precise determination of such quantities, once critical currents obtained within a two-dimensional framework are underestimated.

\Acknowledgement{
LRC, LVT, and ES thank the Brazilian Agency FAPESP for financial 
support, grant numbers 20/03947-2, 19/24618-0, 20/10058-0, 
respectively.}

 \bibliographystyle{elsarticle-num} 
 \bibliography{ref}





\end{document}